\documentclass[prb,twocolumn,superscriptaddress,floatfix,article]{revtex4-1}
\usepackage[english]{babel}
\usepackage[dvips]{graphicx}
\usepackage{color}
\usepackage{latexsym}
\usepackage{amsmath}
\usepackage{amsthm}
\usepackage{amsfonts}
\usepackage{amssymb}
\usepackage{mathrsfs}
\usepackage{natbib}
\usepackage{gensymb}
 
\begin{document}

\title{Tunable ohmic environment  using Josephson junction chains}
\author{Gianluca Rastelli}
\affiliation{Zukunftskolleg \& Fachbereich Physik, Universit{\"a}t Konstanz, 78457 Konstanz, Germany}
\author{Ioan M. Pop}
\affiliation{Physikalisches Institut, Karlsruhe Institute of Technology, 76131 Karlsruhe, Germany}

\begin{abstract}
We propose a scheme to implement a tunable, wide frequency-band dissipative environment  
using a double chain of Josephson junctions.
The two parallel chains consist of identical SQUIDs, with magnetic-flux tunable inductance, 
coupled to each other at each node via a capacitance much larger than the junction capacitance. 
Thanks to this capacitive coupling, the system sustains electromagnetic modes with  
a wide frequency dispersion.
The internal quality factor of the modes is maintained as high as possible, and the damping is introduced    
by a uniform coupling of the modes to a transmission line, itself connected to an amplification and readout 
circuit.
For sufficiently long chains, containing several thousands of junctions, 
the resulting admittance is a smooth function versus frequency in the microwave domain, 
and its effective dissipation can be continuously monitored by recording the emitted radiation in the 
transmission line. 
We show that by varying in-situ the SQUIDs' inductance, the double chain can
operate as tunable ohmic resistor in a frequency band spanning up to one 
GHz, with a resistance that can be swept through values comparable to the resistance 
quantum $R_q= h/(4e^2)\simeq 6.5 \, \mbox{k}\Omega$. 
We argue that the circuit complexity is within reach using current Josephson junction technology.
\end{abstract}

\date{\today}

\maketitle


%
%
%
%
%
%
%
\section{Introduction}
Dissipation in radio-frequency (rf) superconducting quantum electronic circuits \cite{Clerk:2010dh,Devoret:2013jz,Vool:2017hm}
is usually detrimental, giving rise to quantum decoherence. 
However, this does not necessarily have to be the case.
Remarkably, in the last decade, engineered 
dissipation \cite{Kraus:2008jd,Diehl:2008ha,Verstraete:2009kc,Weimer:2010ez} 
played an increasingly prominent  role  in quantum states 
stabilization\cite{Gerry:1993fy,Garraway:1994de,deMatosFilho:1996ds,Poyatos:1996va,Toth:2017dm}   
or even in error correction  schemes \cite{Leghtas:2013kh,Leghtas:2015kd,Wang:2016bt}.

So far, low-impedance dissipative environments have dominated the scene, as they are ubiquitously 
present in rf circuits and can be tailored using standard microwave design strategies.
Designing high impedance environments, with an impedance comparable 
to the resistance quantum $R_q= h/(4e^2) \simeq 6.5 \, \mbox{k}\Omega$, has proven   
more challenging.  
Recently, significant success has been achieved in the fabrication of low-loss high impedance 
environments in the form of 
superinductors \cite{Manucharyan:2009fo,Masluk:2012ib,Bell:2012,Pop:2014gg,Grunhaupt:2018}
or metamaterials \cite{Stockklauser:2017bq,Zhang:2017fs}.
However,  the implementation of wide frequency-band, high-impedance ohmic environments remains an 
unsolved problem.

Tunable, high-impedance ohmic environments are potentially interesting for several applications 
in the field of superconducting electronics.

For instance, quantum simulations of fundamental models to study dissipative quantum phase transitions  
require the exploration of extended regions in their phase diagrams  \cite{Leggett:1987,Weiss:2012}. 
In a single Josephson junction, dissipation leads to a phase transition with suppression of the quantum tunneling of the superconducting phase when 
the effective resistance shunting the junction is swept through the resistance quantum $R_q$.
The phase diagram of such a transition was experimentally explored using different shunting resistances \cite{Penttila:1999kb,Penttila:2000fu}.
In circuit QED, the ratio between the characteristic impedance $Z_c$ of a one-dimensional microwave waveguide and 
the quantum resistance $R_q$ plays the role of the effective fine structure constant between the artificial atoms, 
viz. superconducting qubits, and the electromagnetic field \cite{Devoret:2007gia}, namely  ${\alpha}_{eff} =(Z_c/Z_{vac}) \alpha 
\sim  Z_{c}/R_q$,
with the $Z_{vac}$ the impedance of the vacuum and $\alpha \simeq 1/137$.

Ultra-strong coupling regime in circuit QED has been achieved in experiments in resonant cavities  \cite{Niemczyk:2010gv,Yoshihara:2016bi}
and in open microwave waveguides \cite{FornDiaz:2016bo}, using  galvanic coupling, which is characterized by a dual scaling of 
the coupling strength in matter-radiation interaction\cite{Devoret:2007gia,Peropadre:2013}, e.g. $\sim 1/\alpha_{eff}$. 
This regime  was also obtained experimentally in the effective rotating frame of a driven qubit coupled to  
an LC resonator \cite{Braumuller:2017kd}, with a theoretical extension to an ensemble of resonators  \cite{Leppäkangas:2017}.
Hence, circuits QED designs offer another realization of the spin-boson model, a reference model in the theory of quantum dissipation.
For instance, a recent experiment investigated  transmons coupled to  transmission lines with different coupling 
strength \cite{Magazzu:2017}.
Another  recent approach is based on the use of 1D arrays of Josephson junctions to design the resonant modes of the 
electromagnetic  environment\cite{Martinez:2018,Goldstein:2013kq}.
In these systems it is desirable to have the ability to controllably sweep the relevant parameter over a wide range, 
i.e. the strength of the dissipative interaction between the quantum system and 
its environment \cite{Yagi:1997gc,vanderWal:2003fh,Jones:2013gn}. 
Varying {\sl in situ} the resistance opens the route for addressing novel issues as, for instance, quenching 
in the dissipative phase transition by varying rapidly the external dissipation across the critical point.

This class of environments could also be an asset for quantum state preparation and   
stabilization \cite{Pastawski:2011ga,Tuorila:2017he}  
and autonomous quantum error correction via bath 
engineering \cite{Mirrahimi:2014js,Cohen:2014gy}.
For example, in the context of coherent cat states preparation,
tuning the dissipative strength and the characteristic impedance might provide a significant resource 
\cite{Mirrahimi:2014js,Cohen:2014gy,Cohen:2016}.

%
%
%
%
\begin{figure}[t!]
\includegraphics[scale=0.6,angle=0.]{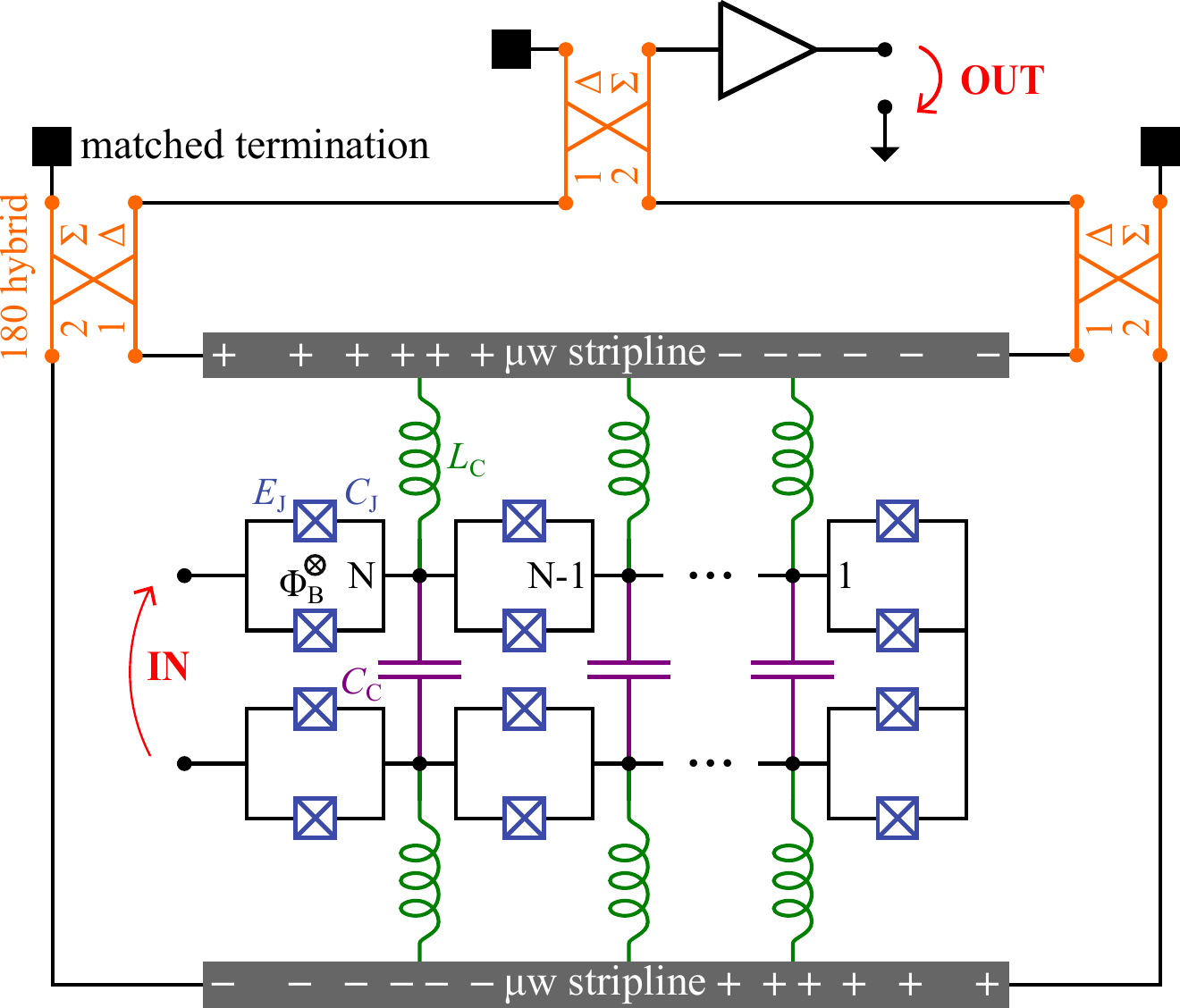}
\caption{
Schematic of the two parallel Josephson junction chains (PJJC). 
The effective Josephson junctions are implemented using SQUIDs 
threaded by a magnetic flux $\Phi_B$ to achieve a tunable Josephson inductance 
$L_J= \Phi_0^2/[ 8 \pi^2 E_J \cos(2\pi\Phi_B/\Phi_0) ] $, where $\Phi_0=h/(2e)$ is the flux quantum.
The two chains are coupled via the capacitances $C_C$, with $C_C \gg C_J$.
Each node is inductively coupled through the inductance $L_C$  to a balanced microstrip transmission line. 
The transmission line is connected via $180^{\circ}$ hybrid couplers to a standard coaxial line, 
ideally without reflections, and the signal can then be routed to a microwave amplification and readout chain.
}
\label{fig:schema_system}
\end{figure}
%
%
%

In this work we analyse the possibility to realize a tunable high-impedance environment, ohmic in a wide frequency range, 
using two coupled parallel Josephson junction chains (PJJC), as depicted in Fig.~\ref{fig:schema_system}.

Josephson junctions (JJ) are versatile circuit elements, with widespread use in quantum mesoscopic systems, 
thanks to their intrinsic low dissipation and amenable non-linearity.
They are the building blocks of superconducting quantum bits 
(qubits) \cite{Clarke:2008ud,Martinis:2009br,Wendin:2017bw}, 
hybrid systems \cite{Xiang:2013hma,Pirkkalainen:2013gh},  
or Josephson photonic circuits\cite{Hofheinz:2011jc,Cassidy:2017jg}.
Josephson junction chains exhibit rich and interesting many-body  physical properties\cite{Fazio:2001jv},
which can be influenced relatively accurately by circuit design and fabrication parameters.
They have constituted the platform of choice for the investigation of 
quantum fluctuations of the phase induced by charge interactions,  i.e. quantum phase 
slips \cite{vanderZant:1988bl,Matveev:2002gn,Pop:2010bb,Maibaum:2011da,Pop:2012fi,Manucharyan:2012cz,Rastelli:2013,Susstrunk:2013io,Ergul:2013fr,Garanin:2016,Ergul:2017ff}, 
or quantum fluctuations of the charge induced by Josephson 
tunneling \cite{Hermon:1996jg,Haviland:1996fg,Corlevi:2006jy,Syzranov:2009fz,Homfeld:2011fw,Vogt:2015cf,Cedergren:2015fd, Cedergren:2017}. 

In the phase regime, where the Josephson energy $E_J= \hbar I_c/(2e)$, with $I_c$ the junction critical current, 
dominates over the charging energy of the junction $E_C = e^2/(2C_J)$, with $C_J$ the junction capacitance, 
Josephson junction chains have already been investigated as custom-designed eletromagnetic environments, 
implementing metamaterials  \cite{Rakhmanov:2008,Hutter:2011cj,Zueco:2012,Peropadre:2013do,Masahiko:2015,Iontsev:2016bo},
resonators with  tunable non-linearity \cite{Weissl:2015vm,Muppalla:2017}, or
parametric amplifiers \cite{Eichler:2014dk}. 
The success of many-junction devices in the phase regime  $(E_J \gg E_C)$,  
paves the way towards more complex architectures, such as the two coupled Josephson junction chains  
we propose in Fig.~\ref{fig:schema_system} to implement a tunable, high-impedance ohmic environment.

The PJJC device shown in Fig.~\ref{fig:schema_system} consists of two JJ chains 
capacitively coupled to each other at each node and inductively coupled to 
 a stripline microwave transmission line. 
Each element is formed by a SQUID, with $E_J\gg E_C$, threaded by a magnetic flux $\Phi_B$ that allows tuning of  
the Josephson inductance  $L_J= \Phi_0^2/[ 8 \pi^2 E_J \cos(2\pi\Phi_B/\Phi_0) ]$. 
The coupling capacitance  between the chains $C_C$ are designed to be dominant compared to $C_J$ $(C_J \ll C_C)$, 
which imposes a dense and linear dispersion relation over a wide frequency range.
Dissipation is introduced via the inductive coupling (using $L_C$) of the chains to an on-chip microwave transmission line, 
which is itself connected to an amplification and read-out circuit with using 180$^{\circ}$ hybrid couplers to mode-match 
between the on-chip transmission line and the standard $50$ $\Omega$ coaxial cable. 
This matching is important to avoid the formation of standing waves in the transmission line, 
which would result in a non-uniform coupling of the PJJC eigenmodes to the $50$ $\Omega$ environment.

We show that for sufficiently long chains, with $N$ in the range of $10^{3}$, 
the resulting real part of the impedance at the input port of the PJJC 
is a smooth function versus frequency in a band of $\sim 1$ GHz, and its value 
$\simeq \sqrt{L_J/C_C}$  is tunable in-situ,  straddling the resistance quantum.
Additionally, owing to the fact that dissipation is introduced via  coupling to a transmission line, 
one can continuously monitor the photons emitted by the device of interest, connected to the input port 
of the PJJC.

Notice that, in our proposal, Josephson junctions are only used as linear inductances and could be in principle replaced by 
geometric inductors.
Nevertheless, Josephson inductors are very convenient for this proposal, as they offer three essential ingredients: 
a) an intrinsically lossless medium, 
b) an ultra-compact inductor, much larger than the geometric inductance of an equivalent size wire, and
c) tunability via the Josephson effect, when implemented in the shape of a SQUID.

The paper is structured as follows.
In Sec.~\ref{sec:model}  we discuss the admittance of a single JJ chain.
We compare a  phenomenological model for dissipation, based on an infinite number of dissipationless junctions, 
with models for finite size JJ chains, formed by $N$ dissipative junctions. 
In Sec.~\ref{sec:double-JJchain} we analyze the effective circuit of the PJJC in Fig.~\ref{fig:schema_system}
and show its equivalence to a single chain formed by $N$ JJs.
In Sec.~\ref{sec:disc}, we discuss the realistically achievable values for 
the admittance of the PJJC device, 
taking into account the limited range of experimentally feasible parameters.  
Finally, we draw our conclusions in Sec.~\ref{sec:summary}.

%
%
%
%
%
%
%
\section{Admittance of a single chain formed by N  lumped elements}
\label{sec:model}

In sec.~\ref{sec:model_A}  we recall the emergence of  an ohmic  resistor 
in the mathematical limit of an infinite line formed by dissipationless JJs  
acting as linear inductances and capacitances.    
In sec.~\ref{sec:model_B}  
we demonstrate that a similar result can be obtained for finite chain lengths, N, if 
the JJ element of the chain is intrinsically dissipative. 
In a first example, assuming typical measured values for the intrinsic dissipation of the JJ, 
the real part of the resulting admittance can only become a smooth function vs. 
frequency for chain lengths of the order $N=10^5$. 
In a second example, we engineer the dissipation, and we can obtain a smooth 
admittance vs. frequency for much shorter chains with $N \sim 10^3$. 
The later results will be directly applicable to the PJJC device 
(as shown in Sec.~\ref{sec:double-JJchain}).

%
%
%
\subsection{Ohmic admittance of a dissipationless JJ chain in the thermodynamic limit}
\label{sec:model_A}

%
%
%
%
\begin{figure}[b!]
\includegraphics[scale=0.7,angle=0.]{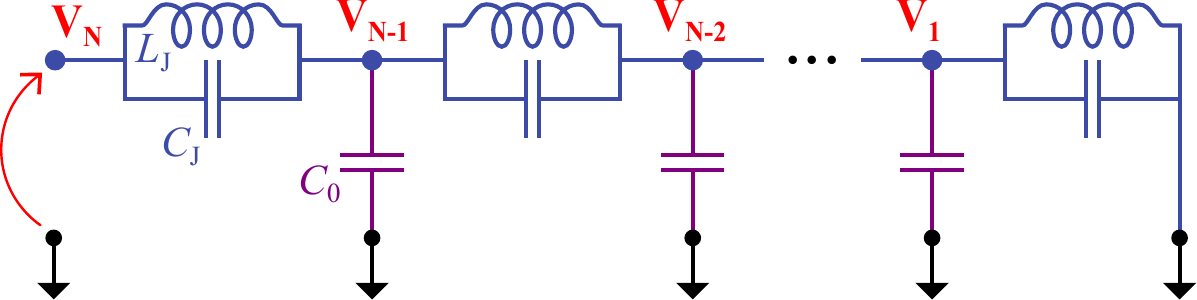}
\caption{Model for a single chain formed by an inductance 
$L_J$ in parallel to a capacitances $C_J$.
Each island is connected by a capacitance $C_0$ to the ground.}
\label{fig:schema_JJchain}
\end{figure}
%
%
%

We consider the chain shown in Fig.~\ref{fig:schema_JJchain} formed by a series of  inductances 
$L_J$, in parallel with capacitances $C_J$,  with $C_0$ connecting each node to the ground. 
Introducing  the two admittances 
$Y_{J}(\omega)  = i \omega C_J + 1/(i \omega L_J)$ and 
$Y_{0}(\omega)  = i \omega C_0$, 
we write  Kirchhoff's laws  for  current conservation at nodes $n=1,\dots,N$, 
in terms of  the voltages $v_n=V_n(\omega)$, in frequency domain  
%
%
%
%
%
%
\begin{equation}
\label{eq:Kirchhoff}
\hspace{-0.3mm}
Y_{J}(\omega) \left(v_{n} - v_{n-1}  \right) \!=\! Y_{J}(\omega) \left( v_{n-1} - v_{n-2} \right)+ Y_{0}(\omega)v_{n} \, ,
\end{equation}
%
%
%
%
%
%
with the boundary condition $V_0=v_0=0$.
We consider a vector of dimension $N-1$ composed of the voltage values at nodes $n=1,\dots,N-1$.
Then,  Eq.~(\ref{eq:Kirchhoff}) can be cast in the following tridiagonal matrix  form 
%
%
%
%
%
%
\begin{equation}
\label{eq:matrix}
\left(
\begin{array}{ccccc}
   a(\omega) 		&     - 1 			&          0				&    \dots		&  \dots			\\
  -1				&        a(\omega)	&      - 1				&     0		&  \dots			\\
    0				&       - 1  			&       a(\omega)		&   -1		&  \dots			\\
    \dots			&      \dots			&    \dots				&    \dots		&  -1			\\
   \dots 			&      \dots			&       0				&  -1			&    a(\omega)  
\end{array}
\right)
\left(
\begin{array}{c}
v_{N-1} \\
v_{N-2} \\
\dots \\
\dots\\
\dots\\
v_{1} 
\end{array}
\right)
=  
\left(
\begin{array}{c}
v_N\\
0\\
\dots \\
\dots\\
\dots\\
0
\end{array}
\right)
\end{equation}
%
%
%
%
%
%
with $a(\omega)=2+Y_0(\omega)/Y_J(\omega)$.

The previous matrix has eigenvalues 
$\lambda_k(\omega) =  2 \left[ 1 - \cos (\pi k/ N) \right] + Y_0(\omega)/Y_J (\omega)$ and   
eigenvectors $e_{k}(n)= \sqrt{2/N} \sin(\pi k \, n /N)$ for $k\in[1,N-1]$ defined on the restricted lattice $n\in[1,N-1]$.
The eigenvectors are orthonormal $\sum_{n=1}^{N-1} e_{k}(n) e_{k'}(n) = \delta_{kk'}$ and satisfy the completeness 
relation $\sum_{k=1}^{N-1} e_{k}(n)e_{k}(m)= \delta_{nm}$.
The matrix  appearing in the left of Eq.~(\ref{eq:matrix}) can be 
written as  $\bar{\bar{Y}} = \bar{\bar{U}} \bar{\bar{D}} {\bar{\bar{U}}}^{-1} $ where $ \bar{\bar{D}} $ is the diagonal matrix 
of the eigenvalues,  and $ \bar{\bar{U}}$ ($\bar{\bar{U}}^{-1}$) is a matrix whose $k$-row (-column)
are the components of the eigenvector $k$.
By writing the inverse of the matrix as   
$\bar{\bar{Y}}^{-1}  = \bar{\bar{U}} \bar{\bar{D}}^{-1} \bar{\bar{U}}^{-1} $, 
one can express the voltage at node $N-1$ as a function of the voltage at  node $N$, namely
$v_{N-1}  = v_N  \sum_{k=1}^{N-1} e_{k}^2(n) / \lambda_k(\omega)$, which reads 
%
%
%
%
%
%
\begin{equation}
\label{eq:V_N-1}
v_{N-1} = 
v_N  
\frac{2 }{N} 
\sum_{k=1}^{N-1} 
\frac{ Y_J(\omega) \sin^2\left( \pi k/N \right) }{Y_0(\omega)  + 2 Y_J(\omega)  \left[ 1 - \cos (\pi k/ N) \right]}  .
\end{equation}
The admittance of the chain is defined by the relation  
%
%
%
%
%
%
\begin{equation}
\label{eq:current_JJ}
I(\omega) = Y_J (\omega)\left( v_N - v_{N-1}\right) \equiv Y_{ch}(\omega) \,  v_N \,.
\end{equation}
Inserting Eq.~(\ref{eq:V_N-1}) into  Eq.~(\ref{eq:current_JJ}), using the  relation
$\sin^2(x) = (1-\cos(x))(1+\cos(x))$, and  the sum 
$\sum_{k=1}^{N-1} [1+\cos(\pi k /N) ]=N-1$, the admittance can be expressed as  
%
%
%
%
%
%
\begin{equation}
\label{eq:admittance_JJ}
Y_{ch}(\omega)\!\!=\!\!  \frac{ Y_J (\omega)}{N} 
\!\left( \! 1 
\!+\! 
\sum_{k=1}^{N-1} 
\! 
\frac{  Y_0 (\omega)  
\left[1+\cos\left(\frac{\pi k}{N}\right) \right]}{Y_0(\omega)  + 2 Y_J(\omega)  \left[ 1 -\cos\left(\frac{\pi k}{N}\right) \right]}
\! \right)
\,.
\end{equation}
We can cast the admittance $Y_{ch}(\omega)$ of 
Eq.~(\ref{eq:admittance_JJ}) as the sum of three terms 
%
%
%
%
%
%
\begin{equation}
\label{eq:admittance_JJ_ideal}
Y_{ch}(\omega) = \frac{1}{i \omega N L_J} + i \omega \tilde{C} + Y_{har}(\omega)  \, . 
\end{equation}
$Y_{ch}(\omega)$ is characterized by an effective inductance $N L_J$
at small frequency and an effective capacitance $\tilde{C}$ at large frequency, given by 
%
%
%
%
%
%
%
\begin{equation}
\label{eq:renormalized_C}
\tilde{C} = \frac{C_J}{N} +  \frac{C_0}{2N} \sum_{k=1}^{N-1}
\frac{\cos^2\left(\frac{\pi k}{2N}\right) }{\sin^2\left(\frac{\pi k}{2N}\right)+C_0/(4C_J)}  .
\end{equation}
This same result was obtainted using a different method  
in previous works \cite{Rastelli:2013,Rastelli:2015hz},  with  different   boundary 
conditions 
\footnote{
In this work the boundary condition is that the chain is connected to the ground at the end, 
so that the wavefunctions of the modes vanishes at this point.
This explains  the difference of a factor 2 inside the $\cos-$ function for   
the eigenfrequencies of the chain  studied in this work respect to the frequency 
of the harmonic modes in a Josephson junction ring  studied in 
Ref.\onlinecite{Rastelli:2015hz} and \onlinecite{Rastelli:2013}
}.
The third term  $Y_{har}(\omega)$ in Eq.~(\ref{eq:admittance_JJ_ideal}) 
is related to the electromagnetic eigenmodes of the chain, whose 
spectrum reads 
\begin{equation}
\label{eq:eigenfrequencies-0}
\omega_k =  \, 
\frac{ 2 \omega_0  \sin\left[\pi k/(2N)\right] } 
{ \sqrt{1 + (4 C_J/C_0) \sin^2\left[\pi k/(2N)\right]  } } \, .
\end{equation}
We introduce the characteristic frequencies of the spectrum  
\begin{equation}
\label{eq:parameters-1} 
\omega_0 = \frac{1}{\sqrt{L_JC_0}} \, , \,\,\,  \omega_J = \frac{1}{\sqrt{L_JC_J}} \, ,  \,\, \,
\omega_{m}=\max_k\{\omega_k\}  \, ,
\end{equation}
corresponding, respectively, to the  frequency scale in the linear regime, 
the plasma frequency of the single JJ, and the maximum frequency of the spectrum,  
given by $\omega_m = 2 \omega_0 /\sqrt{1+4C_J/C_0}$, for $N\gg 1$. 
Using the eigenmodes spectrum, $Y_{har}(\omega)$  can be written as 
%
%
%
%
%
%
\begin{equation}
\label{eq:admittance_modes}
Y_{har}(\omega) = 
\frac{i  2  \omega}{N L_J}  
\sum_{k=1}^{N-1}  
\frac{    
\left( 1 - \frac{\omega_k^2}{\omega_J^2} \right)  
 \left( 1 - \frac{\omega_k^2}{\omega_{m}^2} \right)  
}{\omega_k^2 -\omega^2-2 i \varepsilon \omega_k} \, , 
\end{equation}
In Eq.~(\ref{eq:admittance_modes})  we  added phenomenologically an imaginary part $\varepsilon>0$
in the denominator, which yields a finite real part for the admittance.
From Eq.~(\ref{eq:eigenfrequencies-0}) for the modes, and from their corresponding admittance in 
Eq.(\ref{eq:admittance_modes}), in the limit of $\varepsilon=0$, 
we can recover previous theoretical results \cite{Rastelli:2013,Rastelli:2015hz}.

It is now interesting to discuss the behavior of $Y _{har}(\omega) $ at small frequency, 
with $\pi k /(2N) \ll 1$,  such that we can assume a linear spectrum $\omega_k \simeq  \omega_0 \pi k /N$.
We define $K$ as the approximated fraction of modes in the linear part of the spectrum, 
a number that scales as $K \propto N$.
At low frequency, the numerator of Eq.(\ref{eq:admittance_modes}) 
converges to one $(\omega_k \ll \min[\omega_m,\omega_J])$.
Then, for $\omega>0$ and provided that the imaginary part is much smaller than the lowest eigenfrequency 
$\varepsilon \ll \omega_{k=1}$ (which is equivalent to the requirement that $N \varepsilon = \mbox{const.}$ or 
$K \varepsilon = \mbox{const.}$), one can approximate the real part of the admittance to a sum of Lorentzian functions  
\begin{equation}
\label{eq:appoximated_finite_eps_2}
\mbox{Re}\left[Y _{har} (\omega)\right]  \simeq \frac{1}{Z_0}
\left(  \frac{\pi \omega_0}{N}  \right) \sum_{k=1}^{K} 
\frac{\omega }{\omega_k} \frac{ \varepsilon/\pi }{ {\left( \omega_k - \omega  \right)}^2 +  \varepsilon^2 } \, , 
\end{equation}
with the characteristic impedance of the line 
\begin{equation}
\label{eq:parameters-2} 
Z_0 = \sqrt{  L_J/C_0 } \, .
\end{equation}
Considering the limit of infinite length of the chain $N\rightarrow \infty$  (or equivalently  $K \rightarrow \infty$) 
and keeping constant the product $N \varepsilon$ (or $K \varepsilon$), 
the admittance of the chain, given by Eq.~(\ref{eq:appoximated_finite_eps_2}), 
converge to an ohmic behavior $1/Z_0$.

%
%
%
\subsection{Ohmic admittance of a finite size dissipative JJ chain}
\label{sec:model_B}

In the following we use a different approach compared 
to the previous section to introduce dissipation in the JJ chain. 

We review two types of dissipative 1D JJ chains, where the dissipation can either be 
intrinsically associated to all circuit elements 
(see Fig.~\ref{fig:microscopic-model-1} and section \ref{subsubsec:intrinsic}),
 or it can be added in a controlled manner, 
 in parallel with the ground capacitance $C_0$, using a coupling inductor $L_C$ 
 (see Fig.~\ref{fig:microscopic-model-2} and 
 section \ref{sec:engineered-dissipation}).
In both cases, the chain is composed of the effective junction admittances $\mathbb{Y}_J(\omega)$, 
and the effective admittances to the ground $\mathbb{Y}_0(\omega)$.
Applying  the current conservation at each node, 
similarly to Eq.~(\ref{eq:Kirchhoff}), we obtain 
\begin{equation}
\label{eq:Kirchhoff_1}
\hspace{-1mm}
\mathbb{Y}_{J}(\omega) \left(v_{n} - v_{n-1}  \right)=  
\mathbb{Y}_{J}(\omega) \left( v_{n-1} - v_{n-2} \right)+ 
\mathbb{Y}_0(\omega)v_{n} \, .
\end{equation}
Then one can repeat identically the steps following  Eq.~(\ref{eq:Kirchhoff}) in the previous section 
to obtain   the admittance  
\begin{equation}
\label{eq:admittance_JJ_dissipation}
\mathbb{Y}_{ch}(\omega)\!\!=\!\!  \frac{\mathbb{Y}_J (\omega)}{N} 
\!\left(\! 1 
\!+\!  
\sum_{k=1}^{N-1}
\frac{\mathbb{Y}_J (\omega) \mathbb{Y}_0 (\omega)  
\left[1+\cos\left(\frac{\pi k}{N}\right) \right]}{\mathbb{Y}_0(\omega)  + 2 \mathbb{Y}_J(\omega)  \left[ 1 -\cos\left(\frac{\pi k}{N}\right) \right]}
\!\right) .
\end{equation}
In the next two subsections 
we apply this result to two hypothetic circuit implementations: 
1) the case of intrinsic dissipation associated with any real superconducting circuit element, 
and 2) a particular implementation of engineered dissipation, using resistive on-chip thin-films.
We refer to them  as intrinsic dissipation and engineered dissipation, respectively.

\subsubsection{JJ chain with intrinsic dissipation}
\label{subsubsec:intrinsic}

%
%
%
%
\begin{figure}[b!]
\includegraphics[scale=0.7,angle=0.]{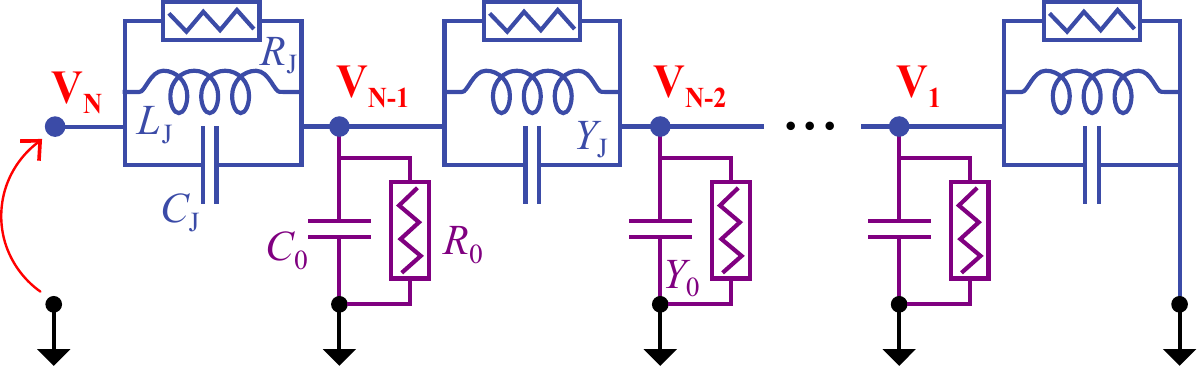}
\caption{
Circuit model for the JJ chain with intrinsic dissipation, 
with shunt resistances $R_J$ in parallel with each junction, 
and $R_0$ in parallel with the capacitance to the ground.}
\label{fig:microscopic-model-1}
\end{figure}
%
%
%

We introduce dissipation by considering the inductances and capacitances to be nonideal elements,
indicated by the resistances $R_J$ and $R_0$ in the circuit model of Fig.~\ref{fig:microscopic-model-1}.
$R_J$ takes into account the finite dissipation in a single JJ, potentially 
associated to (nonequilibrium) quasiparticles above the 
superconducting gap \cite{Vool:2014cw,Wang:2014dv}
or other imperfections of the JJ dielectric barrier \cite{Wang:2015cv}. 
Similarly, $R_0$ accounts for dielectric losses in $C_0$. 
Then we have 
\begin{eqnarray}
\mathbb{Y}_J(\omega) &=& i \omega C_ J + 1/(i \omega L_J) + 1/R_J \, , \\
\mathbb{Y}_0(\omega) &=& i \omega C_0 + 1/R_0 \, .
\end{eqnarray}
Focusing on the limit in which the two shunt resistances  are much larger than 
the characteristic  resistance of the chain, 
$R_J,R_0 \gg Z_0$, 
following a method analogous to the one used in the previous sections, 
one can find the following approximate expression for the 1D JJ chain admittance, 
\begin{equation}
\label{eq:main-00}
\mathbb{Y}_{JJ}^{(0)}(\omega)
\!\simeq\!
\frac{i 2  \omega}{N L_J}  \mathbb{A}^{(0)}(\omega)
\left( 1 - \frac{\omega^2}{\omega_J^2} \right)  
\sum_{k=1}^{N-1}  
\frac{  \left( 1 - \frac{\omega_k^2}{\omega_{m}^2} \right)    
   }{\omega_k^2 -\omega^2-i\omega\eta_k^{(0)}}   , 
\end{equation}
with the spectrum $\omega_k$  given by 
Eqs.~(\ref{eq:eigenfrequencies-0}) and (\ref{eq:parameters-1}). 
As expected, Eq.~(\ref{eq:main-00})  and Eq.~(\ref{eq:admittance_modes}) 
have a similar form  
\footnote{
Eq.~(\ref{eq:main-00}) still contains a part of renormalized, 
high-frequency capacitance, in contrast to Eq.~(\ref{eq:admittance_modes}) which  
includes only the contribution of the harmonic modes.
Since we are interested to the real part in $\mbox{Re}\left[ \mathbb{Y}_{JJ}(\omega) \right]$, 
this difference is not important.
}.
The complex amplitude $ \mathbb{A}^{(0)}(\omega) = 1-i/(\omega R_0 C_0) $ in  Eq.~(\ref{eq:main-00})
reduces to   $ \sim 1$ at  frequency $\omega R_0C_0 \gg 1$, and 
the damping coefficients  for each mode $k$ are given by 
\begin{equation}
\eta_k^{(0)} =
\frac{1}{R_0 C_0} 
+ \left( \frac{1}{R_J C_J}  - \frac{1}{R_0 C_0} \right) 
\frac{\omega_k^2}{\omega_J^2}  \, .
\end{equation}
Notice that the damping is independent of the eigenmode number 
for frequencies $\omega_k \ll \omega_J$.

%
%
%
\begin{figure}[t!]
\includegraphics[scale=0.35,angle=270.]{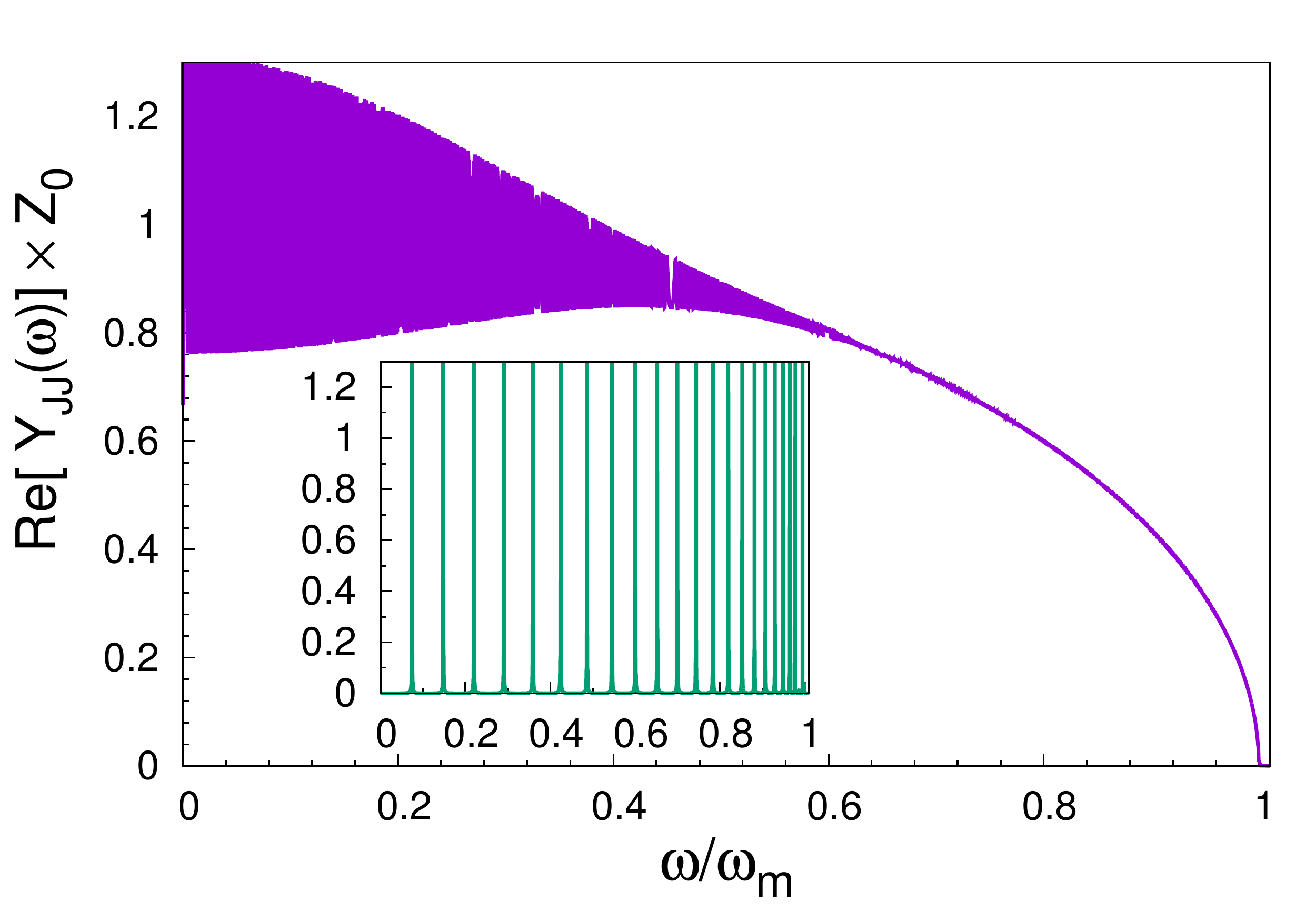}
\caption{The admittance of the 1D JJ chain with intrinsic dissipation (following Eq.~\ref{eq:main-00})
for $N=10^{5}$ and parameters $C_0/C_J=10$,  $R_0=R_J=5 \cdot 10^{4} Z_0$.
For clarity, the inset shows the results for a much shorter chain, with N=25.}
\label{fig:intrinsic-results}
\end{figure}
%
%
%

For $N\gg 1$,  it is expected that the admittance of the system saturates to a smooth function of  frequency. 
The typical length at which the discreteness of the modes disappears is reached when
\begin{equation}
\label{eq:N_c_0}
N \gg \pi R_0/Z_0   \, ,
\end{equation}
i.e. the spacing between the low-frequency modes is much smaller 
than the width of the individual peaks.
Since in typical JJ chains $R_0 \sim 100 \, \mbox{M}\Omega$  \cite{Masluk:2012ib}, 
with a characteristic impedance of the JJ chain $Z_0 \sim \mbox{k}\Omega$,
from Eq.~(\ref{eq:N_c_0})  we get a minimum required number of junctions 
$N \agt 10^{5}$, a  number that is difficult to achieve 
in experimental JJ devices. 

In Fig.~\ref{fig:intrinsic-results} we plot the calculated real part of the JJ chain admittance, 
following Eq.~(\ref{eq:main-00}), for $N=10^5$. 
The inset shows the same calculation for a  short chain with $N=25$, 
to evidence the discrete mode structure of the JJ chain admittance. 
For $N=10^{5}$, the admittance at low frequency still shows large amplitude oscillations 
caused by the discreteness of the eigenmodes spectrum,    
pointing out that even longer chains are needed to  achieve an ohmic behavior in JJ chains 
with intrinsic dissipation.

\subsubsection{JJ chain with engineered  dissipation}
\label{sec:engineered-dissipation}

%
%
%
%
\begin{figure}[b!]
\includegraphics[scale=0.7,angle=0.]{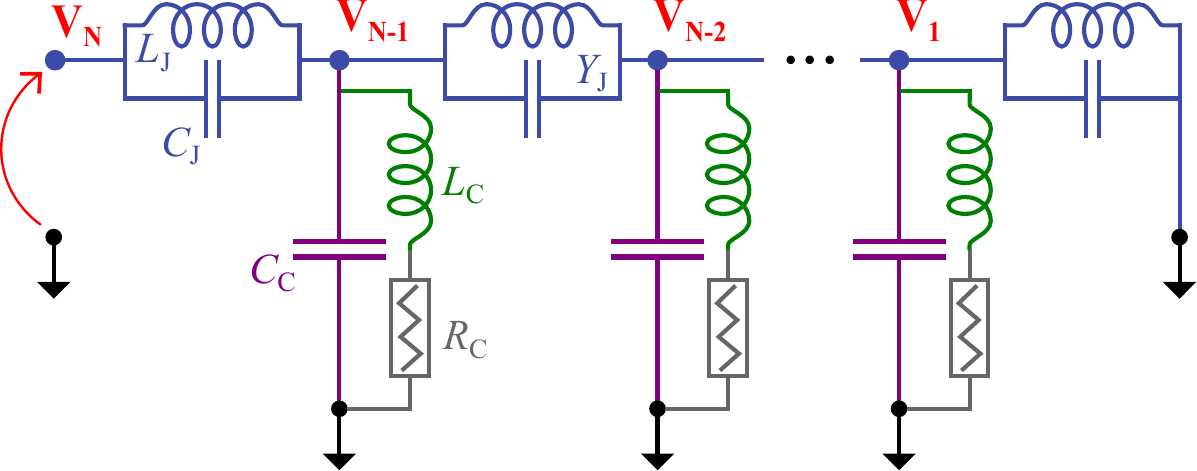}
\caption{
Circuit model for the JJ chain with engineered dissipation. 
The dissipative element $R_C$ 
is introduced using a coupling inductor $L_C$, in parallel with the coupling capacitor $C_C$ 
connected to the ground.}
\label{fig:microscopic-model-2}
\end{figure}
%
%
%

Hereafter we neglect the large intrinsic resistances $R_J$ and $R_0$ associated with 
the dissipative part of non-ideal capacitances and inductances.
As shown in Fig.~\ref{fig:microscopic-model-2}, we  assume $\mathbb{Y}_J(\omega)$ to be 
a pure immaginary admittance,  
 whereas the element $\mathbb{Y}_0(\omega)$ is constructed using  an ideal 
capacitance $C_C$,  in parallel with the series combination $R_C$ and $L_C$.
\begin{eqnarray}
\mathbb{Y}_J(\omega) &=& i \omega C_ J + 1/(i \omega L_J)  \, , \\
\mathbb{Y}_0(\omega) &=& i \omega C_ C  + \frac{1}{R_C+i \omega L_C} \, .
\end{eqnarray}
The inductance $L_C$ opens a gap in the spectrum and the eigenmodes are now given by
\begin{equation}
\label{eq:eigenfrequencies-2}
\Omega_k =   \frac{2}{\sqrt{L_J C_C}}  
\sqrt{ 
\frac{  \sin^2\left[ \pi k/(2N) \right]  + L_J/(4L_C)  } 
{ 1 + (4 C_J/C_C) \sin^2\left[\pi k/(2N)\right]   } 
} \, , 
\end{equation}
with the maximum frequency of the spectrum given by 
$\omega_m = 2  \sqrt{ [1/L_J+1/(4L_C)] / (C_C+4C_J) }$, 
and the minimum frequency  $\omega_{c}=1/\sqrt{L_C C_C}$,
for  $N\gg \max[1,\pi\sqrt{C_J/C_C}]$.  
It is also convenient to introduce the characteristic impedance  
\begin{equation}
\label{eq:Z_C}
Z_C=\sqrt{L_J/C_C} \, .
\end{equation}
Focusing on the frequency range containing the spectrum, 
$\omega_{c} < \omega < \omega_m$, and in  the regime
\begin{equation}
C_J\ll C_C \, , \quad L_J \ll L_C \, , \quad \frac{R_C}{Z_C} \sqrt{\frac{L_J}{L_C}} \ll 1 \, ,
\end{equation}
using the method of Sec.~\ref{sec:model_A}, 
we obtain an approximate expression 
for the real part of the admittance of the chain:  
\begin{equation}
\label{eq:main-0}  
\mathbb{Y}_{JJ}(\omega)
\!\simeq\!
\frac{i 2  \omega}{N L_J}  \mathbb{A}(\omega)
\left( 1 - \frac{\omega^2}{\omega_J^2} \right)  
\sum_{k=1}^{N-1}  
\frac{  \left( 1 - \frac{\Omega_k^2}{\omega_{m}^2} \right)    
   }{\Omega_k^2 -\omega^2-i\omega\eta_k(\omega)} .
\end{equation}
The complex amplitude   $ \mathbb{A}(\omega) $ and the functions   $\eta_k(\omega)$ are now given by 
\begin{eqnarray}
 \mathbb{A}(\omega) &=&  \!\!
 \left(  1 - \frac{\omega^2_c}{\omega^2} -  \frac{i R_C}{\omega L_C} \right) 
 \left[ 
 \frac{1+L_J/(4L_C)}{1-C_JC_C/(L_JL_C)}
 \right] \!, \, \label{eq:factor_A_eng} \\
\eta_k(\omega)  &=&\frac{R_C/L_C}{1-C_JL_J/(C_CL_C)} 
\left(
1-\frac{\Omega_k^2-\omega^2_{c}}{\omega^2}  
\right)\!.  \label{eq:gamma_eng}    
\end{eqnarray}
Notice that the damping coefficients for each mode  are frequency dependent, and they 
increase up to $R_C/L_C$ as the frequency decreases towards the minimum frequency of the spectrum.
As we will show in the next paragraph, the resulting increase of the spectral line-with of the modes at lower frequencies 
allows 
a decrease by two orders of magnitude of the minimum required $N$, 
compared to the previous section.

%
%
%
%
\begin{figure}[t!]
\includegraphics[scale=0.35,angle=270.]{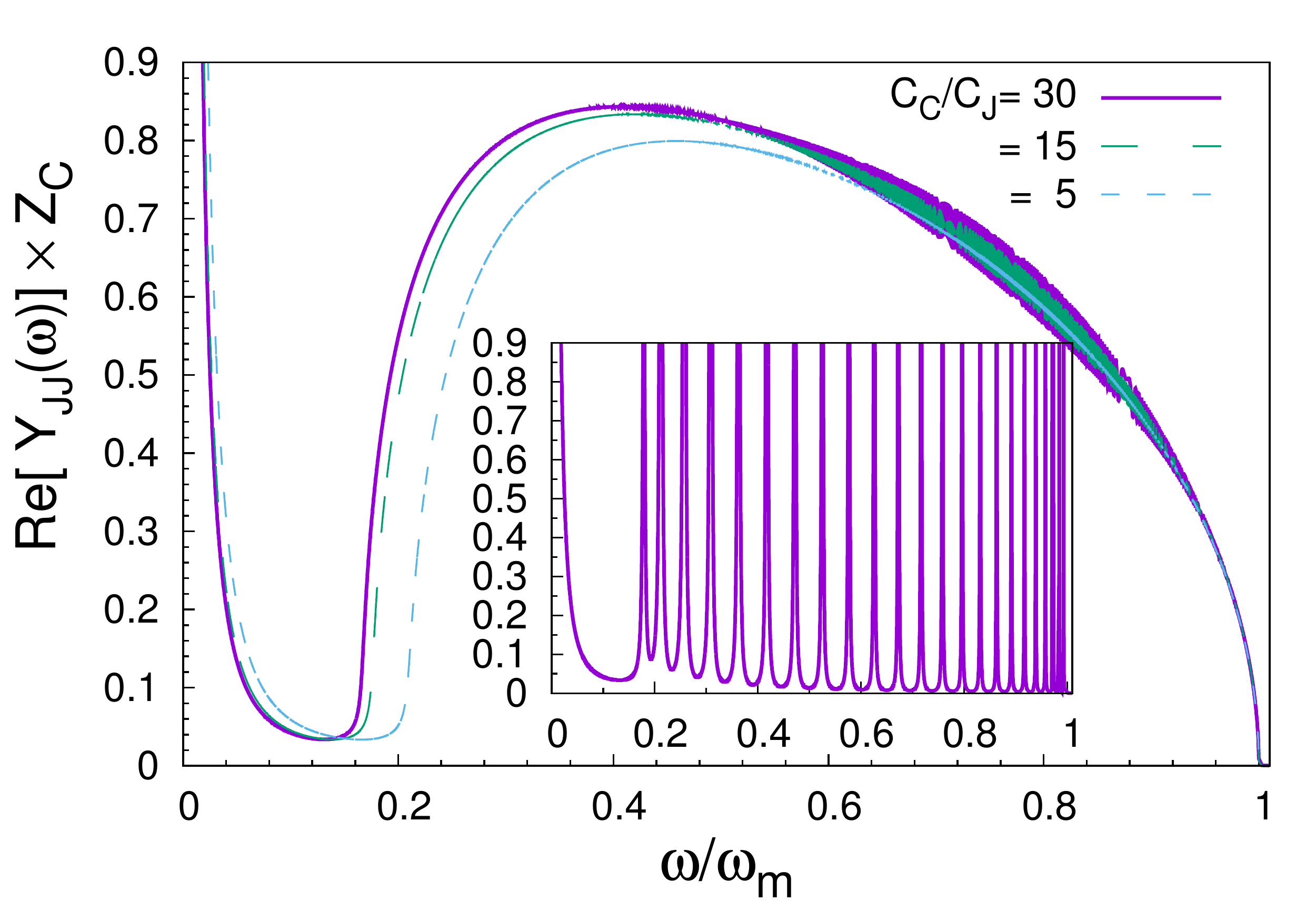}
\caption{
The admittance of the 1D JJ chain with engineered dissipation 
(following Eq. ~\ref{eq:main-0} ),  for $N=5000$ and parameters $L_C/L_J=10$, 
$R_C/Z_C=0.1$, and different ratios $C_C/C_J$. 
For clarity, the inset shows the results for a much shorter chain, with $N=25$, for $C_C/C_J=30$. 
}
\label{fig:results_finite_CJ}
\end{figure}
%
%
%

In the limit $N \gg \pi \sqrt{L_C/L_J}$, 
we can estimate the length above which the discreteness of the modes 
disappears in the admittance 
\begin{equation}
N  \gg \pi {\left( L_C/L_J \right)}^{\frac{3}{4}} \sqrt{Z_C/R_C} \, . 
\end{equation} 
For experimentally feasible parameters such as $R_C=50~\Omega$,  
$Z_C \sim \mbox{k}\Omega$, and $L_C/L_J=10$, the minimum required number of JJs is $N \sim 10^3$. 
Using these parameters,
in Fig.~\ref{fig:results_finite_CJ} 
we plot the calculated real part of the 1D JJ chain admittance according to 
Eq.~(\ref{eq:main-0}), for N=5000.
We observe a  smooth behavior of the admittance 
in a wide frequency range, although some oscillations due to the granularity of the spectrum 
still appear in the high-frequency range. 
These oscillations are a consequence of the fact that the mode damping 
decreases as the frequency approaches the upper edge of the spectrum, see 
Eq.~(\ref{eq:gamma_eng}).

We conclude this section by emphasizing that the introduction of dissipation in a 1D JJ chain via a coupling inductor to the ground (see Fig.~\ref{fig:microscopic-model-2})
allows the design of quasi-ohmic dissipative environments,
 functioning in a relatively wide band. 
The required system parameters, such as chain lengths in the range of $10^3$, 
although ambitious, are not unrealistic for state-of-the-art JJ technology.

%
%
%
%
%
%
%
\section{Admittance of the double chain with engineered dissipation}
\label{sec:double-JJchain}

Following the design of a 1D JJ chain with engineered dissipation introduced in 
Fig.~\ref{fig:microscopic-model-2}, in this section 
we discuss a similar proposal, the PJJC device shown in
Fig.~\ref{fig:schema_system}, 
where dissipation is not added via on-chip dissipative elements, 
like in sec.~\ref{sec:engineered-dissipation}, 
but rather by a uniform coupling to a microwave transmission line, 
which could also allow the continuous monitoring of the dissipated energy.

\begin{figure}[t!]
\includegraphics[scale=0.7,angle=0.]{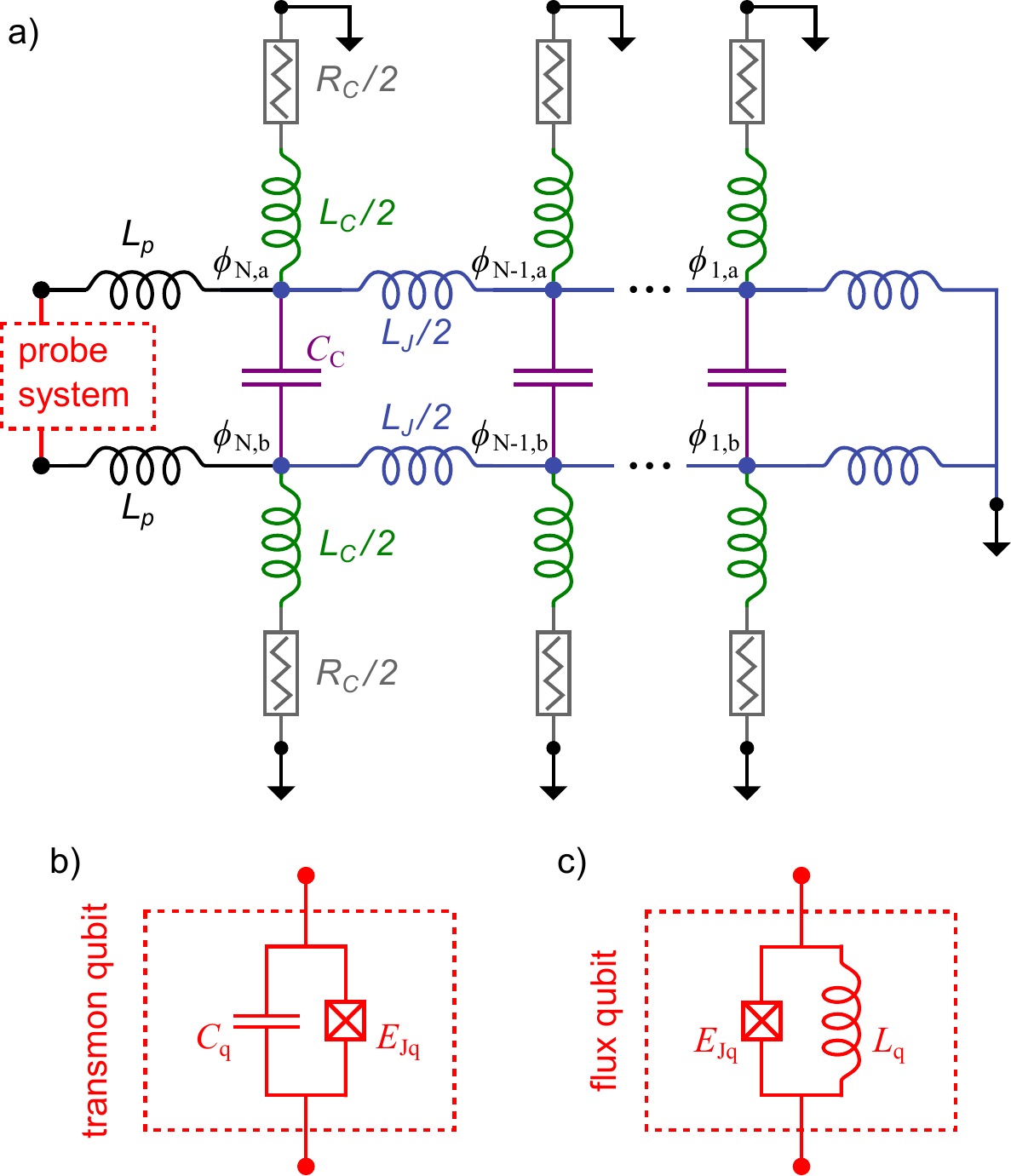}  
\caption{(a) Equivalent circuit of the PJJC  shown in Fig.~\ref{fig:schema_system}. 
The  microstrip transmission line is modelled as a resistance $R_C$.
This floating configuration is convenient to connect the PJJA impedance 
to a probe system such as a transmon (b) or flux (c) qubit. 
}
\label{fig:equivalent-system}    
\end{figure}

We analyze theoretically an equivalent circuit of the PJJC, as shown in  Fig.~\ref{fig:equivalent-system}, 
which captures one essential ingredient of the PJJC proposal, 
namely a uniform dissipation distributed along the nodes 
of the chain.
The resulting PJJA impedance can be connected to a probe system, 
for example a flux \cite{Mooij:1999ex,Chiorescu:2003jm} 
or transmon \cite{Schreier:2008gs} qubit, 
coupled via the inductance $L_P$.  
The microstrip transmission line, which is ideally reflectionless and 
matched to a standard coaxial cable  $(50 ~\Omega)$, acts as a 
resistor $R_C$ at each node of the chain. 
Under the condition of local mirror reflection symmetry for the two chains, 
we shown that the PJJC  is 
equivalent to a single chain connected directly to the 
ground via $C_C$, as shown in Fig.~\ref{fig:microscopic-model-2}. 
%
%
Hereafter, we assume the relevant regime $C_C \gg C_J$ and neglect the junction capacitance $C_J$ 
to simplify the formulas, although the treatment can be  extended to 
the case $C_J \neq 0$. 
Similarly, we consider the local  ground capacitance of each island negligible, 
i.e. $C_0 \ll C_C$.

%
%
%
%
%
%
%

We quantize the circuit of Fig.~\ref{fig:equivalent-system} using the standard 
method to construct the Lagrangian and equations of motion 
for a quantum electromagnetic circuit formed by lumped elements  \cite{Devoret:2004-LesHouches}.
We  use the phase nodes variables $\Phi_{n,s}$, with $n=1,\dots,N$ and $s=a,b$ 
for the two chains connected via the capacitances $C_C$. 
The index $n$ runs from $n=1,\dots,N-1$ for the two chains, with the boundary condition $\Phi_0=0$. 
The index $n=N$  is for the probe system, the  qubit, 
formally described by the node phases $\Phi_{N,a}, \Phi_{N,b}$. 
For $\Phi_{n,s}$, with $n=1,\dots,N-1$, the dynamics of the system is ruled by the equations of motion 
\begin{eqnarray}
C_C \frac{d^2\! \left(  \Phi_{n,a} -\Phi_{n,b} \! \right)}{dt^2} &=&  
-\frac{2}{L_J} \left(2   \Phi_{n,a} -  \Phi_{n-1,a} - \Phi_{n+1,a} \right)  \nonumber \\
&-& \int^{+\infty}_{-\infty}\!\!\!\!\!\!\!\!\!  dt'  \,  2  Y_c(t-t') \frac{d \Phi_{n,a}}{dt'} \, , \label{eq:motion_chain_up} 
\end{eqnarray}
and
\begin{eqnarray}
C_C \frac{d^2\! \left(  \Phi_{n,b} -\Phi_{n,a} \! \right)}{dt^2} &=&  
-\frac{2}{L_J} \left(2   \Phi_{n,b} -  \Phi_{n-1,b} - \Phi_{n+1,b} \right)  \nonumber \\
&-& \int^{+\infty}_{-\infty}\!\!\!\!\!\!\!\!\! dt'  \, 2 Y_c(t-t') \frac{d \Phi_{n,b}}{dt'}  \, , \label{eq:motion_chain_down} 
\end{eqnarray}
with the external admittance   
\begin{equation}
Y_c(t) = \theta(t)  e^{-t/\tau_c} / L_C\, , \quad Y_c(\omega) = 1/(R_C+i\omega L_C) \, , 
\end{equation}
where $\theta(t)$ is the {\sl theta} function,  and  $\tau_c = L_C/R_C$.
It is now  convenient to introduce the phase differences 
\begin{equation}
\phi_n = \Phi_{n,a} - \Phi_{n,b}  \, .
\end{equation}
We can also define equivalently the variables corresponding to the average local phase $\sim \Phi_{n,a}  + \Phi_{n,b} $.
We remark that a finite $C_J$ does not introduce any coupling between the two families of modes. 
Then, taking the difference between the equations of motion
Eqs.~(\ref{eq:motion_chain_up}, \ref{eq:motion_chain_down}),  
we get
\begin{equation}
\label{eq:phi_n_t}
C_C \! \frac{d^2 \phi_n }{dt^2} \! \! =  
\! -\frac{1}{L_J} \left(2   \phi_{n} \!-\!  \phi_{n-1} \!-\! \phi_{n+1} \right)  
\!-\! 
\int^{+\infty}_{-\infty}\!\!\!\!\!\!\!\!\! dt'  Y_c(t-t') \frac{d \phi_{n}}{dt'}  \, .
\end{equation}
We remark that, for a  system characterized by local capacitances $C_C^{(n)}$ and inductances 
$L_J^{(n,a)}$, $L_J^{(n,b)}$, Eq.~(\ref{eq:phi_n_t}) remains 
valid if mirror reflection symmetry is present.
%
%
The set of equations Eq.~(\ref{eq:phi_n_t}), valid for $n=1,\dots,N-1$,  
can be cast in the following matrix form 
\begin{equation}
C_C \! \frac{d^2 \vec{\phi}' }{dt^2} \! \! = 
\! - \frac{ \bar{\bar{M}}_{TB} }{L_J} \vec{\phi}' 
\!-\!
\int^{+\infty}_{-\infty}\!\!\!\!\!\!\!\!\!\! dt'  Y_c(t-t')  \frac{d \vec{\phi}' (t')}{dt'}
+\frac{1}{L_J} \!\!\! 
\left[\!\!
\begin{array}{c}
0 \\
0 \\
\dots \\
\phi_{N}
\end{array}
\!\! \right] \, , 
\end{equation}
with the vector $^{T}\vec{\phi}' = \left( \phi_1, \dots,\phi_n,\dots,\phi_{N-1}\right)$, 
and the tight binding matrix $( \bar{\bar{M}}_{TB} )_{nm} = 2\delta_{nm} - \delta_{n-1,m} - \delta_{n+1,m}$.
%
%
After the unitary transformation  $\theta_k= \sum_{n=1}^{N-1} e_k(n) \phi_n$  that  diagonalizes the 
tight binding matrix $( \bar{\bar{M}}_{TB} )_{nm} $, with eigenvalues and eigenvectors
\begin{equation}
\hspace{-2.mm}
\lambda_k = 2\left[1 - \cos\left(\frac{ \pi k}{ N}\right)\right] \, ,  e_k(n)=\sqrt{\frac{2}{N}}  \sin\left( \frac{\pi k n}{N} \right) \, , 
\end{equation}
for $k=1,\dots,N-1$, we obtain the equation 
\begin{equation}
\label{eq:theta_k}
C_C \frac{d^2 \theta_k}{dt^2} = -\frac{\lambda_k}{L_J} \theta_k 
-\int^{+\infty}_{-\infty}\!\!\!\!\!\!\!\! dt'  \, Y_c(t-t') \frac{d\theta_k }{dt'} 
+ \frac{e_k(N-1)}{L_J} \phi_N \, .
\end{equation}
Notice that the dissipative term 
is not changed after the transformation from the local node variables 
to the harmonic modes of the double chain.
Going in the frequency space via Fourier transform, 
we have as inhomogeneous solution of Eq.~(\ref{eq:theta_k}) 
\begin{equation}
\label{eq:solution_theta_k_omega}
\theta_k(\omega) = \chi_k(\omega) e_k(N-1) \phi_N(\omega) 
\end{equation}
with the dimensionless susceptibility 
\begin{equation}
\label{eq:susceptibility_k}
\chi_k(\omega)  \!=\! \!  
\frac{ \left( \frac{i}{\tau_c}  -  \omega\right)/(L_JC_C) }
{\omega^3 
- \frac{i}{\tau_c} \omega^2 
- \Omega_k^2 \omega 
+ \frac{i}{\tau_c} \left(  \Omega_k^2 -  \omega_c^2 \right)    }
\!\equiv\! 
\sum_{i=1}^{3} \!
\frac{A^{(k)}_{i}}{\omega-z^{(k)}_{i}}
\end{equation}
where the eigenfrequencies of the modes $\Omega_k$ are given by  Eq.~(\ref{eq:eigenfrequencies-2}), 
$z^{(k)}_{i}$ correspond to the roots of the cubic in the denominator of $\chi_k(\omega)$
in Eq.~(\ref{eq:susceptibility_k}), and the factors $A^{(k)}_{i}$ are  given in appendix \ref{app:susceptibility}). 
After some algebra (see appendix \ref{app:susceptibility} for details) 
we can express the solution 
\begin{eqnarray}
\label{eq:solution_theta_k_t}
\theta_k(t) &=& 
-  e_k(N-1)
\int^{+\infty}_{-\infty}\!\!\!\!\!\!\!\!\! dt' \theta(t-t') 
\sum_{i=1}^{3} \frac{ A_i^{(k)} }{z_i^{k}} e^{i z_i^{(k)} (t-t')} 
\frac{d \phi_N}{dt'}   \nonumber \\ 
&+& \frac{e_k(N-1)}{\lambda_k} \phi_N(t) \, .
\end{eqnarray}
Finally, we consider the equation for the node associated to the probe system (the qubit, at node $n=N$)
 in terms  of the phase difference $\phi_N$.
For simplicity, we set $L_P=L_J$ and write 
\begin{eqnarray}
%
%
\frac{d}{ dt} \left( \frac{\partial \mathcal{L}_q}{\partial \dot{\phi}_N}\right)
&=&
\frac{\partial \mathcal{L}_q}{\partial \phi_N}
 -\frac{1}{L_J} \left(\phi_{N} -  \phi_{N-1}  \right)   \nonumber \\
&=&\!\!
\frac{\partial \mathcal{L}_q}{\partial \phi_N}
\!-\! \frac{\phi_{N}}{L_J}  + \frac{1}{L_J} \sum_{k=1}^{N-1} e_{k}(N-1) \theta_k ,
\label{eq:qubit}  
\end{eqnarray}
with  $ \mathcal{L}_q$  the Lagrangian function of  the phase difference of the qubit probe:  
its explicit form is not relevant for  our analysis. 
Inserting the solution Eq.~(\ref{eq:solution_theta_k_t}) into Eq.~(\ref{eq:qubit})  we get 
the equation for the phase difference $\phi_N$ of the qubit
\begin{equation}
\frac{d}{ dt} \left( \frac{\partial \mathcal{L}_q}{\partial \dot{\phi}_N}\right)
\!\!
=
\!
\frac{\partial \mathcal{L}_q}{\partial \phi_N}
-\frac{ \phi_N }{N L_J}   
-   \int^{+\infty}_{-\infty}\!\!\!\!\!\!\!\!\! dt' Y_{ch}(t-t') 
\frac{d \phi_N}{dt'}  , 
\end{equation}
in which we used the property $\sum_{k=1}^{N-1} e^2_{k}(N-1)/\lambda_k = 1-1/N$,  and 
we set the admittance of the double chain to 
\begin{equation}
\label{eq:Y_JJ_time}
Y_{JJ} (t)  =  \frac{ \theta(t) }{L_J} \sum_{k=1}^{N-1} e^2_{k}(N-1) 
\sum_{i=1}^{3} \frac{ A_i^{(k)} }{z_i^{k}} e^{i z_i^{(k)} t }  \, .
\end{equation}
By using some algebraic relations of 
the root $z_{i}^{(k)}$  (see appendix \ref{app:susceptibility} for details), 
we derive the final expression for the admittance  in 
Eq.~(\ref{eq:Y_JJ_time})  in frequency space   
\begin{equation}
\label{eq:main-result_1}
Y_{JJ}(\omega)
\!\!= \!\! 
\frac{i 2  \omega}{N L_J}    \left(  \! 1 \!-\! \frac{\omega^2_c}{\omega^2} \!-\!  \frac{i}{\omega \tau_c} \!  \right) 
\sum_{k=1}^{N-1}  
\frac{\left( 1+ \frac{L_J}{4L_C}   \right)  \left( 1 - \frac{\Omega_k^2}{\omega_{m}^2} \right)}
{\Omega_k^2 -\omega^2-i\omega \gamma_k(\omega)}  \, , 
\end{equation}
with the damping functions 
\begin{equation}
\label{eq:main-result_2}
\gamma_k(\omega)  =\frac{1}{\tau_c} 
\left(
1-\frac{\Omega_k^2-\omega^2_{c}}{\omega^2}  
\right) \, . 
\end{equation}
\mbox{} \\
Equations (\ref{eq:main-result_1}) and (\ref{eq:main-result_2}) 
represent the goal of this section: 
the admittance $Y_{JJ}(\omega)$ 
corresponds exactly to the limit  $C_J/C_C \rightarrow 0$ of the 
admittance  $\mathbb{Y}_{JJ}(\omega)$  in Eq.(\ref{eq:main-0})  
of a single chain with engineered interaction, for $L_J\ll L_C$.

To summarize, we showed that the effective circuit shown in Fig.~\ref{fig:equivalent-system} (case  $C_J\ll C_C$) 
for the PJJC device of Fig.~\ref{fig:schema_system} 
is equivalent to the admittance of the single JJ chain with engineered dissipation  discussed 
in Sec.~\ref{sec:engineered-dissipation} for vanishing junction capacitance $C_J=0$.
Therefore, in the following we will use  Eq.~(\ref{eq:main-0}),(\ref{eq:factor_A_eng}) and 
Eq.~(\ref{eq:gamma_eng}) to calculate the PJCC admittance for circuits with experimentally feasible parameters.

%
%
%
%
\section{PJJC admittance with experimentally feasible parameters} 
\label{sec:disc}

%
%
%
%
\begin{figure}[t!]
\includegraphics[scale=0.31,angle=270.]{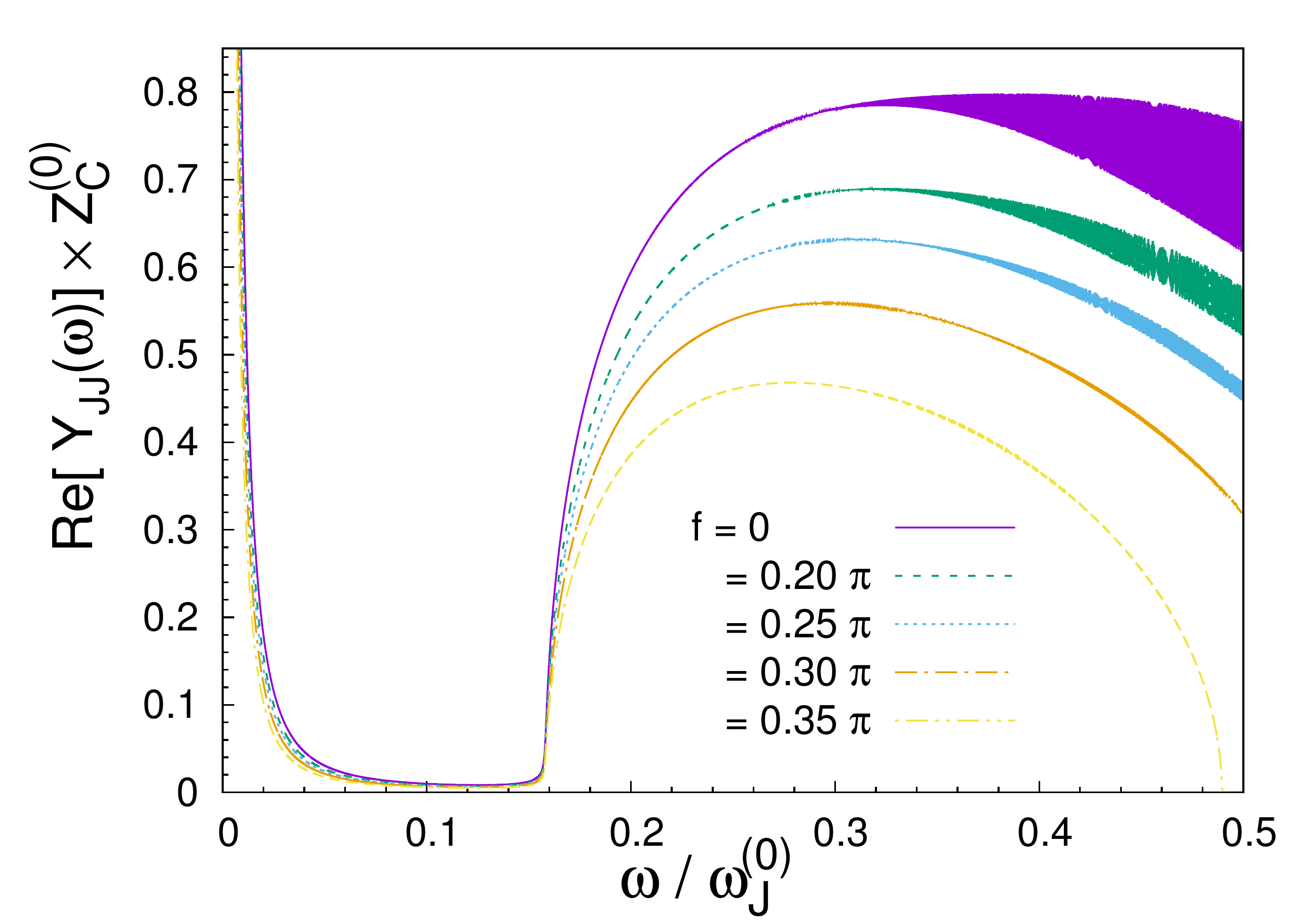}
\caption{
The frequency dependent admittance of the PJJC can be tuned 
in-situ by a perpendicular magnetic field $\Phi_B$, 
threading the SQUID junctions of the PJJC device in Fig.~\ref{fig:schema_system}. 
The reduced flux-bias is defined as $f=2\pi \Phi_B/\Phi_0$. 
The admittance and the frequency are respectively scaled with 
the characteristic impedance  $Z_C^{(0)} = Z_C(0)$, and the plasma 
frequency $\omega_J^{(0)}=\omega_J(0)$ at zero flux $f=0$. 
The PJJC parameters are the following: $N=8000$, $C_J/C_C=0.25$, and at $f=0$ 
the inductance ratio $L_C/L_J^{(0)}=10$ and the resistance ratio $R_C/Z_C^{(0)}=0.025$. 
}
\label{fig:results_flux_1}
\end{figure}
%
%
%

In the PJJC, each Josephson inductance  is tuned by the applied magnetic flux as 
$L_J= L_J^{(0)} / \cos(f) $ with $f= 2\pi\Phi_B/\Phi_0$ the reduced magnetic flux 
and $L_J^{(0)} = \Phi_0^2/(8 \pi^2 E_J)$ the zero-field inductance.  
The plasma frequency of the Josephson  
junctions, as well as the eigenmodes of the chain are also flux tunable.
From Eq.~(\ref{eq:Z_C}) 
it directly follows that the characteristic impedance $Z_C(f)$ 
can be tuned in-situ by the biasing field $\Phi_B$ 
through the SQUID loops composing the PJJC (see Fig.~\ref{fig:schema_system}). 
An example of the scaled admittance of the PJJC for different flux biases 
is reported in Fig.~\ref{fig:results_flux_1}. 
As expected, increasing the Josephson inductance by applying an external magnetic flux 
leads to a decrease of the admittance of the system, accompanied by a slight change in the overall frequency dependence. 
The frequency range in which the PJJC admittance can be considered ohmic reduces with applied flux. 
Depending on the desired application of the PJJC, one can trade by design its flux tunability 
for a wide bandwidth with ohmic behavior, or vice-versa.

\begin{figure}[t!]
\includegraphics[scale=0.31,angle=270.]{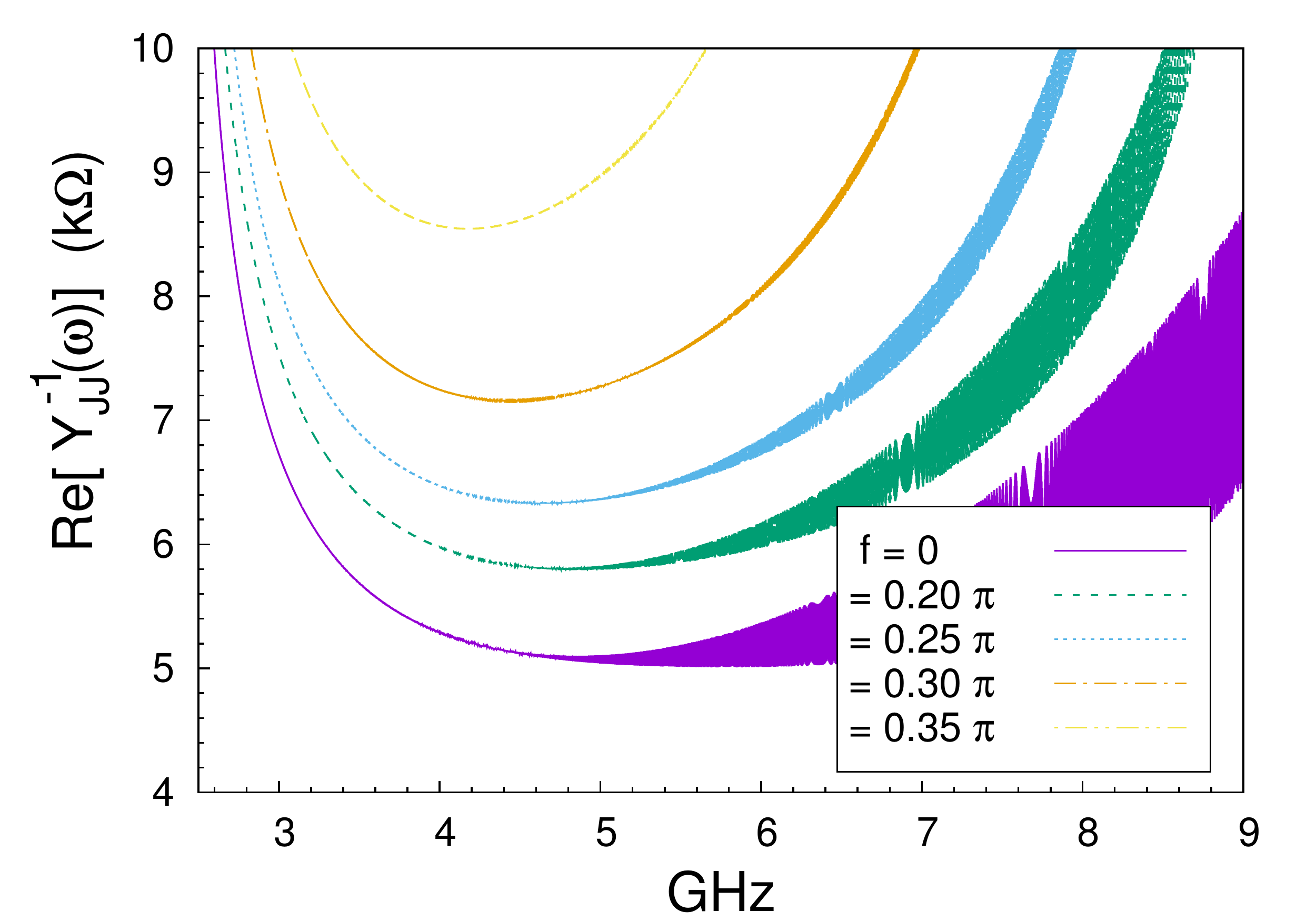}
\caption{
The effective resistance of the PJJC electromagnetic environment shown in Fig.~\ref{fig:schema_system}, 
vs. frequency, for experimentally relevant circuit parameters. 
We chose $R_C=50 \, \Omega$ and, at flux bias $f=0$, the characteristic impedance 
$Z_C^{(0)}= 4 \, \mbox{k}\Omega$, with a  plasma frequency $\omega_J^{(0)} =15 \, \mbox{GHz}$.
The other PJJC parameters are:  $N=8000$, $C_C/C_J=4$ and the inductance ratio $L_C/L_J^{(0)}=10$  at $f=0$.
Notice that as we increase the flux bias $f$, the effective resistance of the environment increases to values above the resistance quantum. 
The value of $f$ can be increased beyond the $0.35 \pi$ threshold shown in the figure, 
which will further increase the effective resistance of the PJJC device. 
However, the frequency range where the resistance can be considered ohmic 
will continue to decrease, while the chain will become increasingly  non-linear. }
\label{fig:results_flux_2}
\end{figure}

In Fig.~\ref{fig:results_flux_2} we show an example of the PJJC impedance 
for a circuit with experimentally feasible parameters, 
with a characteristic impedance 
$Z_C^{(0)}= 4 \, \mbox{k}\Omega$ and 
the plasma frequency $\omega_J^{(0)} =15 \, \mbox{GHz}$   at $f=0$.
The Josephson inductance at zero flux is $L_J=86 \, \mbox{nH}$, 
the junction capacitance $C_J=1.3 \, \mbox{fF}$, and the coupling capacitance $C_C=5.2 \, \mbox{fF}$. 
The required coupling inductance $L_C=10 L_J$ is in the superinductance regime \cite{Manucharyan:2009fo}.  
It can be implemented either using an array of JJs  \cite{Masluk:2012ib,Bell:2012}
or a high kinetic inductance thin film, such as granular aluminum, or niobium and titanium 
nitrides \cite{Rotzinger:2017jd,Annunziata:2010el,Samkharadze:2016ie,Vissers:2010fk}. 
Below $5$ GHz, the curves appear flat in a range of $\sim 1$ GHz at  $f=0$,   
and $\sim 0.5$ GHz for flux bias $f=0.35 \pi$.
In this frequency range, the impedance of the PJJC can be tuned by the biasing magnetic field  from 
$\sim 5 \, \mbox{k} \Omega$ up to $\sim  9 \, \mbox{k} \Omega$.

The single photon nonlinearity introduced by the JJs can be estimated based on Ref.~\onlinecite{Weissl:2015vm} 
to  be  in the range of $100 \, \mbox{kHz}$.   
This value is orders of magnitude lower than the line-width of the PJJC modes 
(see inset of Fig.~\ref{fig:results_finite_CJ}),  and it can be ignored for low power applications.

%
%
%
%
%
%
%

%
%
%
%
%
%
%
\section{Summary}
\label{sec:summary}

We have demonstrated that the parallel Josephson junction chain device shown in 
Fig.~\ref{fig:schema_system} 
can implement a tunable ohmic environment, 
over a frequency-band of the order of GHz, 
with an effective resistance that can be tuned through the resistance quantum 
$R_q = 6.5 \, \mbox{k}\Omega$. 
The PJJC can be connected to any two-terminal device under test, such as a superconducting qubit or a resonator, 
and its dissipation can be continuously monitored using a low-noise rf amplification chain. 

The PJJC principle of operation can also be applied for constituent SQUIDs with different geometries, 
such as the ones proposed in Ref.~[\onlinecite{Bell:2012},\onlinecite{Zhang:2017fs}], 
implementing even higher impedances and resulting in larger effective resistances.
It is also worth mentioning that the rapid increase of the PJJC resistance at low frequencies (see Fig.~\ref{fig:results_flux_2}) 
protects the device from low energy thermal excitations.

We believe that the tunable, high-impedance ohmic environment implemented by the PJJC 
will be a useful instrument in the route towards quantum simulations of dissipative phase transitions, 
or the engineering of environments for autonomous quantum error correction schemes.

\acknowledgments 
We acknowledge Denis Basko for interesting discussions and useful comments.
GR acknowledges support from by the German Excellence Initiative through the Zukunftskolleg 
of the University of Konstanz,  the DFG through the  SFB 767 and grant RA 2810/1-1, 
and  the MWK-RiSC program, project No.13971016. 
IMP acknowledges support from the Alexander von Humboldt foundation 
in the framework of a Sofja Kovalevskaja award endowed by the German 
Federal Ministry of Education and Research. 
 
\appendix

\section{Susceptibility for the $k$ harmonic modes}
\label{app:susceptibility}

In this section, we list the main steps of the calculations leading to the main results Eq.~(\ref{eq:main-result_1}) 
and Eq.~(\ref{eq:main-result_2}),  in Sec.~ \ref{sec:double-JJchain}, for the admittance of the double shown in 
Fig.~\ref{fig:equivalent-system}. 

The solution for the dynamics of the harmonic modes connected to the   probe  qubit, 
in frequency space is given by Eq.~(\ref{eq:solution_theta_k_omega}) with the susceptibility 
for the single eigenmode $k$ defined by Eq.~(\ref{eq:susceptibility_k}) and with factors given by 
\begin{equation}
A^{(k)}_{1} =  - \frac{ (z^{(k)}_{1} -  \frac{i}{\tau_c}) / (L_JC_C)}
{  (z^{(k)}_{1}-z^{(k)}_{2}) (z^{(k)}_{1}-z^{(k)}_{3} ) } \, ,
\end{equation}
and similar definitions of $A^{(k)}_{2},A^{(k)}_{3}$.
The roots of the cubic satisfy Veta's relations
\begin{eqnarray}
z^{(k)}_{1}  + z^{(k)}_{2} + z^{(k)}_{3}&=& i/\tau_c \label{eq:Veta_1} \\
z^{(k)}_{1} z^{(k)}_{2} + z^{(k)}_{2} z^{(k)}_{3} + z^{(k)}_{1}  z^{(k)}_{3} &=& - \Omega_k^2  \label{eq:Veta_2}  \\
z^{(k)}_{1} z^{(k)}_{2} z^{(k)}_{3} &=&  (i/\tau_c)  \left(  \omega_c^2- \Omega_k^2  \right)     \label{eq:Veta_3} 
\end{eqnarray}
We also have  the sum rules  
\begin{equation}
\label{eq:sum_roots_1}
\sum_{i=1}^{3} \frac{A^{(k)}_{i}}{z^{(k)}_{i}} = \frac{1}{\omega_c^2-\Omega_k^2} \, , 
\end{equation}
\begin{equation}
\label{eq:sum_roots_2}
\sum_{i=1}^{3}\frac{ A^{(k)}_{i}}{z^{(k)}_{i} (\omega-z^{(k)}_{i}) } 
=
\frac{ 1 + \frac{ \omega^2  -i \omega/\tau_c - \Omega_k^2 
}{\Omega_k^2-\omega_c^2}
}
{\omega^3 - \frac{i}{\tau_c} \omega^2  -  \omega_k^2 \omega + \frac{i}{\tau_c} \left(  \omega_k^2 -  \omega_c^2 \right)    } \, .
\end{equation}
We consider the solution in time which  reads 
\begin{eqnarray}
& & \theta_k(t)   =  \nonumber \\
&=&  \alpha_k    \sum_{i=1}^{3} i A_i^{(k)}  \int^{t}_{-\infty}\!\!\!\!\!\!\! dt'   e^{i z_i^{(k)} (t-t' )} \phi_N(t')    \nonumber \\
&=& 
\alpha_k  \sum_{i=1}^{3} \frac{ A_i^{(k)}  }{z_i^{(k)}}  \left[ \! -  \phi_N(t)   
\!+\!\!\!
\int^{t}_{-\infty}\!\!\!\!\!\!\! dt'  
 e^{i z_i^{(k)} (t-t' )}   \frac{d \phi_N(t')}{dt'}   \! \right] 
\end{eqnarray}
in which we  set $ \alpha_k =  e_{k}(N-1) =\sqrt{2/N} \sin[ \pi k  (N-1)/ N]$  and 
 we have used the fact that 
$\lim_{\Delta t \rightarrow -\infty} e^{i z_i^{(k)} \Delta t} \phi_N(t+\Delta t) = 0$ since the 
roots have positive imaginary parts. 
The weighted sum  over modes reduces to 
\begin{eqnarray}
& & \sum_{k=1}^{N-1} \alpha_k \theta_k(t)    
= - \left( \sum_{k=1}^{N-1} \alpha_k ^2  \sum_{i=1}^{3} \frac{ A_i^{(k)}  }{z_i^{(k)}} \right)   \phi_N(t)  \nonumber \\  
&+&
\sum_{k=1}^{N-1} \alpha_k ^2 
\sum_{i=1}^{3}  \frac{A_i^{(k)} }{z_i^{(k)}} 
 \int^{t}_{-\infty}\!\!\!\!\!\!\! dt'   e^{i z_i^{(k)} (t-t' )}  \dot{\phi}_N(t') \nonumber \\
 &=&  \frac{N-1}{N}  \phi_N(t) 
 +  L_J \int^{+\infty}_{-\infty}\!\!\!\!\!\!\! dt' Y_{JJ}(t-t') \frac{d \phi_N(t')}{dt'}  \, , 
\end{eqnarray}\\
with the admittance of the chain given by
\begin{equation}
Y_{JJ} (t)  =  \frac{ \theta(t) }{L_J} \sum_{k=1}^{N-1} \alpha^2_{k} 
\sum_{i=1}^{3} \frac{ A_i^{(k)} }{z_i^{k}} e^{i z_i^{(k)} (t-t')}  \, .
\end{equation}
In frequency domain, the admittance reads  
\begin{equation}
Y_{JJ}(\omega) = - \frac{i}{L_J} 
\sum_{k=1}^{N-1} \alpha_k ^2 
\sum_{i=1}^{3}  \frac{A_i^{(k)} }{z_i^{(k)}(\omega - z_i^{(k)})}
\end{equation}
Using   Veta's relations Eqs.(\ref{eq:Veta_1},\ref{eq:Veta_2},\ref{eq:Veta_3})  
and the sum rules Eqs.(\ref{eq:sum_roots_1},\ref{eq:sum_roots_1}), we 
can obtatin the main results Eq.~(\ref{eq:main-result_1}) 
and Eq.~(\ref{eq:main-result_2}).


%
%
%
\bibliography{tunanle-R_JJchain-references}
%
%
%

\end{document}